\documentclass{appolb}
\usepackage{epsfig}
\begin{document}

\title{GROUND STATE MAGNETIC MOMENTS OF MIRROR NUCLEI STUDIED AT NSCL
\thanks{Presented at the Zakopane Conference on Nuclear Physics,
September 1-7, 2008}}
\author{
P.F. Mantica$^{a,b}$ and K. Minamisono$^{a}$
\address{
$^{a}$ National Superconducting Cyclotron Laboratory,
Michigan State University, East Lansing, Michigan 48824 USA\\
$^{b}$ Department of Chemistry, Michigan State University, 
East Lansing, Michigan 48824 USA }
}
\maketitle

\begin{abstract}
Progress in the measurement of the ground state magnetic moments of 
mirror nuclei at NSCL is presented.  The systematic trend 
of the spin expectation value $<s>$ and the linear behavior
of $\gamma _p$ versus $\gamma _n$, both extracted from the magnetic moments 
of mirror partners, are updated to include all available data.
\end{abstract}


\section{Introduction}

The ground state magnetic dipole moment has sensitivity to the 
orbital and spin components of the state wavefunction, and hence serves 
as an important observable in the study of nuclear structure.  
In particular, the simultaneous consideration
of the magnetic dipole moments of mirror nuclei can provide 
a framework to test present day nuclear structure models.\\

Sugimoto \cite{sug1964} showed that if isospin is a 
good quantum number, the nuclear magnetic dipole moment
could be decomposed into isoscalar and isovector components

\begin{equation}
\mu = \langle \sum \mu _0 (i) \rangle _{J} +
\langle \sum \mu _3 (i) \rangle _{J}
\label{eq1}
\end{equation}

\noindent
where the sum of the isoscalar $\mu _0$ and isovector $\mu _{3}$
moments are taken over all nucleons and $< \mu > _J$ denotes
the expectation value of $\mu$ for the state $M=J$, where $M$ and 
$J$ are the magnetic quantum number and nuclear spin, respectively. 
The isoscalar magnetic moment represents the sum of the 
magnetic moments of the mirror partners

\begin{equation}
(2T+1) \langle \sum \mu _0 (i) \rangle =
\sum_{T_z} \mu (J,T,T_z).
\label{eq2}
\end{equation}

\noindent
Here $T$ is the total isospin and $T_z = (N-Z)/2$.  The 
left-hand side of Eq.\ \ref{eq2} can also be expressed in
terms of the isoscalar spin expectation value $<s>$

\begin{equation}
\langle \sum \mu _0 (i) \rangle =
\frac{J}{2} + 0.38<s>
\label{eq3}
\end{equation}

\noindent
where

\begin{equation}
<s> = \langle \sum s _z (i) \rangle _J
\label{eq4}
\end{equation}

\noindent
and the constant 0.38 is the sum of the magnetic moments of the 
bare proton and neutron.  

The extreme single-particle limit gives $<s> = 0.5$ for odd-$A$
mirror partners whose odd nucleon occupies a single-particle 
orbital with $j = \ell + s$, where $\ell$ is the 
orbital angular momentum. The value $<s> = -J/[2(J+1)]$ results
for the cases when the odd nucleon resides in an orbital with
$j = \ell - s$.  Experimentally deduced $<s>$ values 
generally fall within single-particle expectations, except
for a few instances that are discussed in detail later.

Buck and Perez \cite{buc1983} analyzed the magnetic moments of mirror 
nuclei in a different approach.  They showed that   
a plot of the gyromagnetic ratio, $\gamma =\mu /J$, 
of the odd proton member of the 
mirror pair $\gamma _p$ as a function of the gyromagnetic ratio
of the odd neutron member $\gamma _n$ resulted 
in a straight line.  Further scrutiny of this 
linear dependence of $\gamma _{p}$ on $\gamma _{n}$, 
provided simple expressions for the slope $\alpha$ 
and intercept $\beta$:

\begin{equation}
\alpha = \frac{G_p - g_p}{G_n - g_n}, \qquad
\beta = g_p - \alpha g_n
\label{eq5}
\end{equation}  

\noindent
where $G_x$ and $g_x$ are the 
spin and orbital contributions to the 
$g$-factor, respectively, with $x= p,n$ for
protons and neutrons, respectively.  The extreme single-particle
model gives $\alpha = -1.199$ and $\beta = 1.000$,
while the most recent evaluation of mirror magnetic
moments for $T=1/2$ nuclei by Buck, Merchant,
and Perez \cite{buc2001} produced $\alpha = -1.148\pm0.010$
and $\beta = 1.052 \pm 0.016$.  The small deviation
of the experimental moments from the extreme single
particle expectation was taken to possibly reflect 
meson exchange currents and/or small 
contributions to $\gamma$ from the even nucleon \cite{buc1983}.

Ground state magnetic moment measurements of
the neutron-deficient nuclei $^{9}$C \cite{huh1998}, 
$^{32}$Cl \cite{rog2000}, 
$^{35}$K \cite{mer2006}, and $^{57}$Cu \cite{min2006} 
have been completed at NSCL.  The results for the
odd-$A$ nuclei $^{35}$K ($T=3/2$) and $^{57}$Cu 
$(T=1/2)$ have significantly extended the 
evaluation of $<s>$ and $\gamma _p$ vs.\ 
$\gamma _n$ to heavier masses.  In this paper, the 
experimental approach to magnetic moment measurements at 
NSCL is described, followed by a summary discussion of the 
new magnetic moment values for $^{35}$K and $^{57}$Cu
and the resulting systematic trends of mirror moments at 
higher mass numbers.    

\section{Magnetic moment measurements at NSCL}

Ground state magnetic moments are measured at NSCL using
the technique of nuclear magnetic resonance on $\beta$-emitting
nuclei ($\beta$-NMR).  Nuclei of interest are produced
by bombarding a fixed target with intermediate energy projectiles
from the NSCL coupled cyclotrons.
The incoming beam is made incident on the target at a 
small angle relative to the normal beam direction to break
the reaction plane symmetry and produce a spin-polarized 
secondary beam of high-velocity ions.  The ion species are 
mass separated in the A1900 fragment separator \cite{mor2003},
with the separator tuned to maximize both 
the purity and transmission of the desired radioactive isotope.
An adjustable slit system located at the A1900 intermediate image is 
used to select a portion of the momentum distribution of the 
desired isotopes, which is then transmitted to the 
$\beta$-NMR endstation.

\subsection{Spin polarization}

The production of spin polarized nuclei in intermediate-energy 
heavy-ion reactions was first demonstrated by 
Asahi {\it et al.} \cite{asa1990}, and has been used 
extensively to measure ground state nuclear moments of 
short-lived isotopes at 
RIKEN, GANIL, GSI, and MSU.  A classical treatment of the 
mechanism to describe the nuclear polarization in
such reactions \cite{asa1990} considered conservation 
of linear and angular momentum.  The treatment was 
extended by Okuno {\it et al.} \cite{oku1994} to
account for varying initial reaction conditions.  
Although good qualitative agreement with
experimental measurements  was achieved,
the magnitude of the observed polarization was typically 
a factor of three smaller than predictions.

The extension of ground state magnetic moments of mirror nuclei
to heavier masses at NSCL was enabled by the establishment
of spin polarization in intermediate-energy heavy-ion reactions
where a single nucleon is picked up from the target by
the fast-moving projectile.  The initial measurements of 
Groh {\it et al.} \cite{gro2003} showed that large, positive
spin polarization is obtained near the peak of the momentum
distribution for proton pickup reactions.  Subsequent systematic 
measurements by Turzo {\it et al.} \cite{tur2006} 
at GANIL demonstrated the method
for neutron pickup as well. 

A more accurate prediction of the spin polarization 
realized in intermediate-energy heavy-ion reactions, 
both for nucleon removal and pickup, has been developed
\cite{gro2007}.  Starting with the 
classical kinematic picture discussed above, a Monte 
Carlo simulation that included the addition 
of a more realistic angular distribution of the outgoing
fragments, deorientation 
caused by $\gamma$-ray emission, and corrections 
for the out-of-reaction plane acceptance, was shown to 
reproduce, both qualitatively and quantitatively, the 
polarization observed in intermediate-energy reactions.
The development of an accurate simulation of the 
spin polarization process significantly aided
the execution of the magnetic moment measurements of 
$^{35}$K and $^{57}$Cu described below.

\subsection{$\beta$-NMR method}

The $\beta$-NMR system at NSCL consists of a small 
electromagnet, $\beta$ detectors, and a radiofrequency
(rf) system \cite{man1997}.  The room temperature 
electromagnet has a pole 
gap of 10~cm and operates at a maximum field of 0.5~T.
The $\beta$ detectors consist of two plastic $\Delta E - E$ 
telescopes placed around the sample holder in ``up''
and ``down'' positions relative to the orientation of 
the magnetic field of the electromagnet.  The thin telescope 
element is 0.3-cm thick BC400 plastic
scintillator of dimensional area 
4.4 $\times$ 4.4~cm, while the thick telescope
element is also BC400 scintillator with thickness 
2.5~cm and area 5.1 $\times$ 5.1~cm.
The scintillators are mounted on BC800 plastic light guides that 
allow placement of the 12-stage photomultiplier tubes outside
the fringing field of the electromagnet.  
A vacuum chamber can be placed in the pole gap of the electromagnet
between the two $\beta$ detector telescopes.  The part of the 
vacuum chamber above and below the sample holder
has been removed and replaced with thin plastic 
to reduce the attenuation of $\beta$ particles.
The downstream plate of the vacuum chamber is used to 
mount the sample holder, rf coils, a beam
collimator, and a
solid-state Si detector for monitoring the secondary beam.

The rf system at NSCL was recently upgraded
to allow the simultaneous application of multiple frequencies
to the sample without significant loss of power \cite{min2008}.   
The frequency-modulated signal from one of up to six frequency 
generators can be selected by a double-balanced mixer, 
amplified to a maximum 250 W, and delivered to a high-power
rf box containing a bank of vacuum variable capacitors
and an impedance matching element (either a 50 $\Omega$
resistor or multi-turn transformer).  These two elements,
along with the rf coil surrounding the sample holder, make
up an $LCR$ resonance circuit with a resonance $Q$ factor
of $\sim 20$.  Tuning of the $LCR$ for a given 
frequency is accomplished by setting the variable capacitors
via remote controlled stepper motor units.  Any combination
of capacitors can be selected by way of fast switching relays.
The time sequence for applying each frequency and the 
corresponding capacitance, as well as any necessary
secondary beam pulsing, is controlled by a pulse pattern generator.

\section{Magnetic moment of the drip-line nucleus $^{35}$K}

The proton separation energy of $^{35}$K is only 80~keV 
based on the 2003 Atomic Mass Evaluation \cite{aud2003},
and its vicinity to the proton drip line may reveal new 
and interesting nuclear structure features that may be 
reflected in the ground state magnetic dipole moment.  The 
$^{35}$K ions were produced starting with a primary
beam of $^{36}$Ar at 150~MeV/nucleon via a 
single proton pickup from a Be target followed by
two neutron removal.  The $^{35}$K production rate was  
$~30$~pps/pnA of primary beam.  The NMR was scanned 
between 520 and 620~kHz based on a previous measurement \cite{sch1998}.
A resonance signal was observed at frequency $600 \pm 10$~kHz,
corresponding to a magnetic moment 
$|\mu (^{35}K)| = 0.392 \pm 0.007 \; \mu _N$
\cite{mer2006}.

\section{Magnetic moment of $^{57}$Cu}

The systematic variation of the ground state magnetic moments
of the odd-$A$ Cu isotopes, where neutrons are filling the 
$pf$ shell, are quenched relative to shell model expectations 
\cite{gol2004}.  In addition, the level structure \cite{yur2006,joh2006}
and transition probabilities \cite{yur2004} give a disparate picture 
of the robustness of the $^{56}$Ni double shell closure.
The magnetic moment of the one-proton particle nucleus $^{57}$Cu 
is expected to be $2.5 \; \mu _N$ based on shell models calculations
completed in the full $pf$ shell \cite{hon2004}.  $^{57}$Cu ions
were produced by impinging a $^{58}$Ni primary beam of 
energy 140 MeV/nucleon on a Be target.  The single-proton 
pickup reaction and subsequent two neutron removal 
resulted in a $^{57}$Cu production rate of $350$ pps/pnA 
of primary beam.  A broad NMR scan was completed between
1400 and 2800~kHz, and a resonance signal was observed 
at $2050 \pm 50$~kHz.  The deduced magnetic moment value 
$|\mu (^{57}Cu)| = 2.00 \pm 0.05 \; \mu _N$ \cite{min2006} was 
nearly 20\% smaller than shell model expectations. 

\section{Isoscalar spin expectation values}

\begin{figure}
\epsfig{file=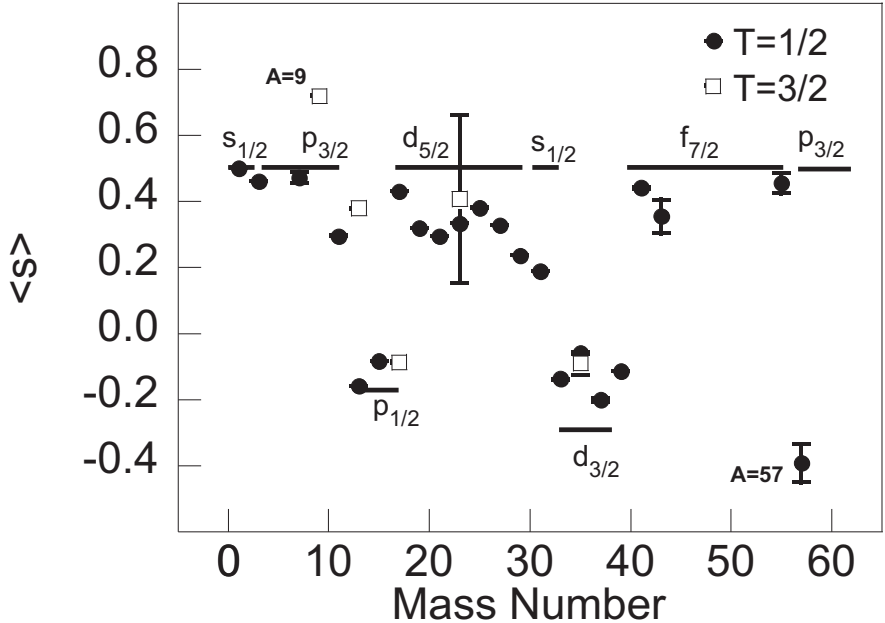,width=5.0in}
\caption{Isoscalar spin expectation values for mirror magnetic moments.
Filled circles are $<s>$ values deduced for $T=1/2$ nuclei, while
the open squares are those deduced for nuclei with $T=3/2$.  The 
limits for $<s>$ from the extreme single-particle model are 
shown by the solid lines.}
\label{fig1}
\end{figure}

The new ground state magnetic moments of $^{35}$K and 
$^{57}$Cu dramatically extend the mass range of known mirror partners
for $T=1/2$ and $T=3/2$ systems.  The systematic trend 
in $<s>$ for odd-$A$ mirror nuclei is depicted in 
Fig.\ \ref{fig1}.  Data are taken from Ref.\ \cite{sto2005} 
with the exception of  $^{35}$K and $^{57}$Cu discussed here and
$^{23}$Al \cite{oza2006} and $^{55}$Ni \cite{pin2009},
for which signs of $\mu$ are taken from theoretical 
predictions.  
Nearly all $<s>$ values are bounded
by the extreme single-particle limits, including the 
new value $<s> = -0.142 \pm 0.020$ deduced for the $A = 35, T = 3/2$
mirror partners, which includes $^{35}$K and $^{35}$S.
It was surprising that the $A = 35, T=3/2$ system followed the 
trends established by more well-bounded nuclei, given that 
$^{35}$K exhibits such a small proton binding energy.  

The two mirror systems whose $<s>$ value lies outside the extreme 
single-particle expectation are those with $A=9, T=3/2$ and 
$A=57, T=1/2$.  The former disparity 
has been linked to possible proton intruder
configurations in the ground state wavefunction of 
$^{9}$C \cite{uts2004}.  The later has been attributed 
to a breaking of the $^{56}$Ni double-magic core \cite{min2006}.

\section{Buck-Perez analysis}

\begin{figure}
\epsfig{file=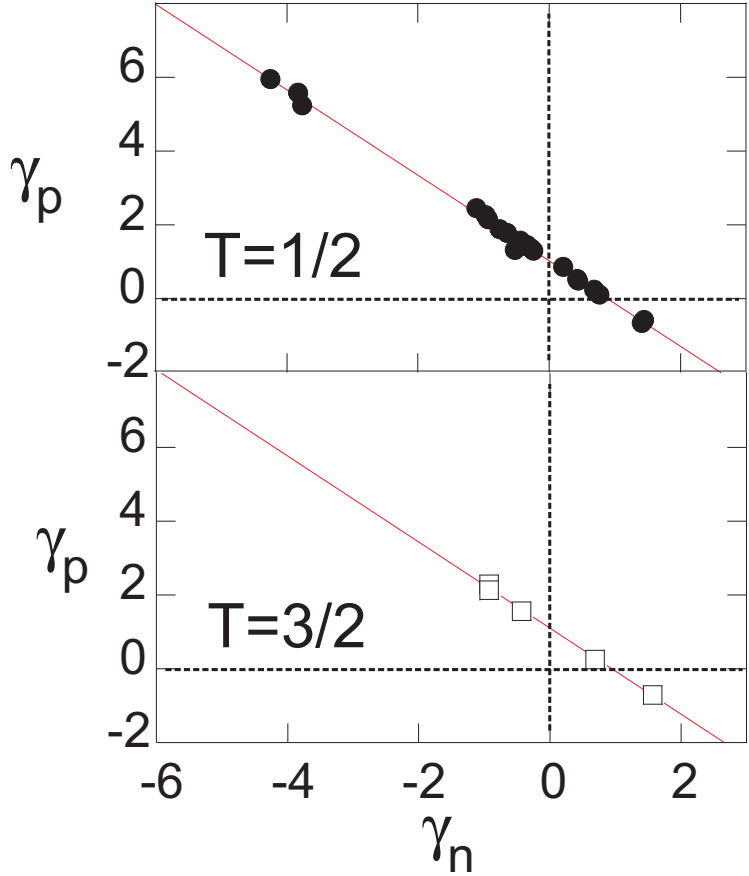,height=5.0in}
\caption{Plot of gyromagnetic ratios for mirror magnetic moments.
Filled circles are for $T=1/2$ nuclei, while
the open squares are those for nuclei with $T=3/2$.
The $\alpha$ and $\beta$ values deduced from the linear fits 
shown as the solid lines in the figure are discussed in the text.}
\label{fig2}
\end{figure}

While the linear behavior of $\gamma _{p}$ vs.\ $\gamma _{n}$ was demonstrated
for $T=1/2$ nuclei, no such analysis had been performed for $T=3/2$
nuclei due to the limited experimental data for magnetic moments 
of $T_z = -3/2$ nuclei near the proton drip line.  There
are now five $T=3/2$ mirror pairs whose ground state magnetic
moments are known, including the $A=35, T=3/2$ system discussed here.  
The $\gamma _p$ vs.\ $\gamma _n$ plots for the 
$T=1/2$ and $T=3/2$ mirror nuclei are presented in Fig.\ \ref{fig2}.

A linear fit for all $T=3/2$ data shown in Fig.\ \ref{fig2}
results in slope $\alpha = -1.165 \pm 0.038$ and intercept
$\beta = +1.101 \pm 0.037$.  The linear trend in 
$\gamma _p$ vs.\ $\gamma _n$ for the $T=3/2$
mirror moments follows that already noted for the $T=1/2$ data.
In addition, the deduced $\alpha$ and $\beta$ values for 
the $T=3/2$ partners 
agree within errors of the values obtain by Buck {\it et al.}
\cite{buc2001} for the $T=1/2$ mirror pairs.

Recently, Perez {\it et al.} \cite{per2008} extended the treatment 
of the linear behavior of mirror magnetic moments by using
shell model estimates to make small modifications to 
$\gamma _p$ and $\gamma _n$.  The inclusion of contributions
of total spin and angular momentum from both odd and even
nucleon types improved the fit to the linear correlations
between gryomagnetic ratios of mirror partners, and 
demonstrated the consistent treatment of 
$\gamma _p$ vs.\ $\gamma _n$ for both 
$T=1/2$ and $T=3/2$ mirror pairs.

We note that the two mirror systems 
at $A = 9, T=3/2$ and $A=57, T=1/2$ that show
anomalous behavior in $<s>$ as shown in Fig.\ \ref{fig1}
follow the linear correlation in $\gamma _p$ vs.\ $\gamma _n$
demonstrated for other known mirror partners.  
The slope $\alpha$ in the Buck-Perez relation 
represents a ratio of the neutron and proton 
spin and orbital $g$-factors, effectively
canceling any systematic spin dependence from the mirror
partners, as opposed to relation for $<s>$, which 
will amplify such spin effects.  The underlying 
case for the disparate behavior of the $A = 9, T=3/2$
and $A = 57, T = 1/2$ mirror moments may lie in the 
spin contribution from low-$\ell$ proton orbitals
that comprise some part of the ground state wavefunction
of the $T_z = -T$ nucleus. 

\section{Future initiatives at NSCL}

We plan to continue our efforts to extend the measurements
of known ground state magnetic moments to heavier nuclei. 
To this end, we are developing a new laser polarizer 
beam line \cite{min2008A} that will receive short-lived, 
low-energy rare isotope beams from the 
gas catcher system at NSCL.  This new experimental 
set up will be operational in 2011 and will 
enable $\beta$-NMR measurements on
refractory elements currently inaccessible at 
ISOL facilities, where collinear laser spectroscopy and
laser polarization measurements have been a staple for 
many years.  

\section{Acknowledgments}

The experimental program on magnetic moments at NSCL has benefitted 
over the years from contributions by T.J.~Mertzimekis, 
A.E.~Stuchbery, D.E.~Groh, A.D.~Davies, M.Huhta, 
J.S.~Berryman, H.L.~Crawford, R.R.~Weerasiri, J.B.~Stoker, 
W.F.~Rogers, G.~Georgiev, D.A.~Anthony, 
and M.~Hass.  This work was supported in part by the U.S.~National 
Science Foundation grant PHY-06-06007.

\end{document}